# MHD stagnation point flow toward a linearly-stretching thermally-insulated sheet with induced magnetic field

Tarek M. A. El-Mistikawy

Dept. Eng. Math. Phys., Faculty of Engineering, Cairo University, Giza 12211, Egypt

**Abstract**

The equations governing the magnetohydrodynamic stagnation point flow toward a non-conducting, thermally insulated, nonporous, linearly stretching sheet are cast in a self similar form. Consistent boundary conditions on the velocity, magnetic field and temperature are invoked. The flow problem involves three parameters- the magnetic Prandtl number, the magnetic interaction number, and the ratio of the stretching rate to the strength of the stagnation point flow. The energy equation includes viscous dissipation and Joule heating, and introduces the Prandtl number as a fourth parameter. Numerical solutions are obtained and sample results are presented.

## 1. Introduction

Articles dealing with the magnetohydrodynamic (MHD) stagnation point flow (SPF) toward a linearly stretching sheet [1-6] introduce simplifying assumptions, in order to cast the problem in a self similar form, that is easily amenable to numerical solutions.

The assumption of high Reynolds number $R_e$ leads to a boundary layer formulation. Only the leading order terms are retained. Higher order effects such as lateral pressure variations within the boundary layer, and the viscous correction to the farfield flow are neglected.

The assumption of small magnetic Reynolds number $R_m$ leads to the neglect of the induced magnetic field. An exception is the work of Ali et al. [7], who included the induced magnetic field within the boundary layer treatment.

Neglect of heat dissipation and Joule heating in the boundary layer heat equation is also a common practice. It allows for self similar formulation when the surface temperature or heat flux is proportional to a power of the streamwise distance from the stagnation point. The exceptions, which include both heat generation mechanisms under the assumption of small $R_m$, are the works of Pop et al. [8] and Nandy [9].

In the present article, it is intended to abandon the abovementioned assumptions, and to show that the problem governing the MHD stagnation point flow toward a linearly stretching sheet can still be cast in a self-similar form. Deeper insight into this problem is gained and benchmark results can be produced for evaluation of numerical methods.

## 2. The flow problem

An electrically conducting incompressible Newtonian fluid is in a two-dimensional SPF toward a linearly stretching, non-conducting, nonporous, thermally insulated sheet.

The equations governing the velocity components ($u$, $v$), the magnetic field components ($r$, $s$) in the ($x$, $y$) directions, respectively, and the pressure $p$ are

$$u_x + v_y = 0 \tag{1a}$$
$$\rho(uu_x + vu_y) + p_x = \rho v(u_{xx} + u_{yy}) - \sigma(us - vr)s \tag{1b}$$
$$\rho(uv_x + vv_y) + p_y = \rho v(v_{xx} + v_{yy}) + \sigma(us - vr)r \tag{1c}$$
$$s_x - r_y = \sigma\mu(us - vr) \tag{1d}$$
$$r_x + s_y = 0 \tag{1e}$$

where $\rho$, $v$, $\sigma$ and $\mu$ are the density, kinematic viscosity, electric conductivity, and magnetic permeability, respectively, all assumed constant.

In the farfield, the electric current density should vanish demanding that the magnetic field be parallel to the velocity field; i.e. $us - vr \sim 0$ as $y \sim \infty$. The farfield problem reduces to a laterally-shifted inviscid SPF problem. Thus, as $y \sim \infty$,

$$u \sim Uax/L \tag{2a}$$
$$v \sim -Ua(y-d)/L \tag{2b}$$
$$p \sim p_d - \tfrac{1}{2}\rho U^2 a^2 [x^2 + (y-d)^2]/L^2 \tag{2c}$$

where $a$ is a measure of the strength of the SPF, and $d$ is a viscously-imposed lateral shift, while $U$ and $L$ are suitably chosen characteristic speed and length, respectively. Note that $p_d$ is not the pressure at the nominal stagnation point $(0,0)$, but, rather, at the viscously-shifted stagnation point $(0,d)$.

The (parallel to the velocity) magnetic field, as $y \sim \infty$, has components

$$r \sim Bx/L \tag{2d}$$
$$s \sim -B(y-d)/L \tag{2e}$$

where $B$ is a measure of the strength of the magnetic field, imposed and induced. The part $Bd/L$ of $s$ is induced in correspondence to the viscous correction $Uad/L$ to $v$.

At the surface $y = 0$, we have the adherence conditions

$$u = Ubx/L \qquad (3a)$$
$$v = 0 \qquad (3b)$$

where $b$ is a measure of the stretching rate.

The problem admits the similarity transformations

$$y = L\eta,\ v = -Uf(\eta),\ u = \frac{U}{L}xf',\ s = -Bg(\eta),\ r = \frac{B}{L}xg'$$

$$p = p_d - \rho U^2 [R_e^{-1}\{f' - a\} + \tfrac{1}{2}f^2 + \tfrac{1}{2}\frac{x^2}{L^2}\{a^2 + R_m^{-1}\beta(g'^2 - 1)\}]$$

where primes denote differentiation with respect to $\eta$, while $R_e \equiv UL/\nu$, $R_m \equiv \sigma\mu UL$, and $\beta \equiv \sigma B^2 L/\rho U$ are the Reynolds number, magnetic Reynolds number, and magnetic interaction number, respectively.

The coupled problem for $f(\eta)$ and $g(\eta)$ is

$$R_e^{-1}f''' + ff'' - f'^2 + a^2 = \beta R_m^{-1}(g''g - g'^2 + 1) \qquad (4a)$$
$$g'' = R_m(gf' - g'f) \qquad (4b)$$
$$f'(0) = b,\ f(0) = 0,\ f'(\infty) = a \qquad (4c,d,e)$$
$$g'(\infty) = 1,\ f(\infty) = ag(\infty) \qquad (4f,g)$$

Condition (4g) results from combining the conditions $f(\infty) = a(\eta - \delta)$ and $g(\infty) = \eta - \delta$, where $\delta = d/L$.

Ali et al. [7] formulated a problem involving Eqs. (4a-f) on the basis of the boundary layer approximation and an imposed magnetic field in the $x$-direction. Instead of condition (4g), they adopted the surface condition $g(0) = 0$, which is a mathematical requirement of symmetry on both sides of an infinitely thin sheet [10]. As such a sheet is physically unfeasible, moreover, symmetry may not be the case, this condition is abandoned in favor of the physically-sound condition (4g).

The magnetic field at the surface $B[(x/L)g'(0), -g(0)]$ is transmitted to the sheet, where the induced part of $B$ is ascertained.

### 3. The thermal problem

The energy equation for the temperature $T$, including viscous dissipation and Joule heating, writes

$$\rho c(uT_x + vT_y) = k(T_{xx} + T_{yy}) + \rho v[2(u_x^2 + v_y^2) + (u_y + v_x)^2] + \sigma(su - rv)^2 \tag{5a}$$

where the specific heat $c$ and the thermal conductivity $k$ are taken to be constant.

The surface condition for a thermally insulated surface is

$$y = 0: \frac{\partial T}{\partial y} = 0 \tag{5b}$$

Toward the farfield, the temperature $T$ cannot tend to a freestream value $T_\infty$. The dissipation term $4\rho v(aU/L)^2$, there, demands a viscous correction to $T_\infty$, so that

$$y \sim \infty: T \sim T_\infty - \frac{4vaU}{cL}\ln(\frac{y-d}{L}) \tag{5c}$$

The similarity expression for $T$ takes the form

$$T = T_\infty + \frac{U^2}{c}[\theta_0(\eta) + \frac{x^2}{L^2}\theta_2(\eta)] \tag{6}$$

leading to the following problems

$$P_r^{-1}R_e^{-1}\theta_2'' + f\theta_2' - 2f'\theta_2 + R_e^{-1}f''^2 + \beta R_m^{-2}g''^2 = 0 \tag{7a}$$

$$\theta_2'(0) = 0, \quad \theta_2(\infty) = 0 \tag{7b,c}$$

$$P_r^{-1}R_e^{-1}\theta_0'' + f\theta_0' + P_r^{-1}R_e^{-1}2\theta_2 + 4R_e^{-1}f'^2 = 0 \tag{8a}$$

$$\theta_0'(0) = 0, \quad \theta_0(\infty) = -4aR_e^{-1}\ln(\eta - \delta) \tag{8b,c}$$

where $P_r \equiv \rho vc/k$ is the Prandtl number. Note that the viscous dissipation is represented by the $f''^2$ term in Eq. (7a) and the $f'^2$ term in Eq. (8a). Joule heating affects $\theta_2$ directly through the $g''^2$ term in Eq. (7a) and affects $\theta_0$ indirectly through the $\theta_2$ term in Eq. (8a).

## 4. Numerical method

Since a closed form solution is not possible, we seek an iterative numerical solution for the flow problem. In the $n^{th}$ iteration, we solve, for $f_n(\eta)$, Eq. (4a) with its right hand side evaluated using the previous iteration solutions $f_{n-1}(\eta)$ and $g_{n-1}(\eta)$, together with conditions (4c,d,e). Then we solve, for $g_n(\eta)$, Eq. (4b) with the known $f_n(\eta)$, together with conditions (4f,g). The iterations continue until the maximum error in $f(\eta_\infty)$, $f''(0)$, $g(0)$ and $g'(0)$ becomes less than $10^{-10}$. For the first iteration, we zero the right hand side of Eq. (4a), which corresponds to $f_0 = a\eta$ and $g_0 = \eta$.

The numerical solution of the problems for $f_n(\eta)$ and $g_n(\eta)$ utilizes Keller's two point, second order accurate, finite-difference scheme [8]. A uniform step size $\Delta\eta = 0.01$ is used on a finite domain $0 \leq \eta \leq \eta_\infty$. The value of $\eta_\infty$ is chosen sufficiently large in order to insure the asymptotic satisfaction of the farfield conditions. The non-linear terms in the problem for $f_n(\eta)$ are quasi-linearized, and an iterative procedure is implemented; terminating when the maximum error in $f_n(\eta_\infty)$ and $f_n''(0)$ becomes less than $10^{-10}$.

Having determined $f(\eta)$ and $g(\eta)$, we solve the problems for $\theta_2(\eta)$ then for $\theta_0(\eta)$, using Keller's scheme on the same grid. However, the farfield condition (8c) on $\theta_0$ poses a problem which is solved by the transformation $\theta_0 = -4aR_e^{-1}\ln(\phi)$ leading to the following problem for $\phi(\eta)$

$$P_r^{-1}R_e^{-1}\phi'' - P_r^{-1}R_e^{-1}\phi'^2/\phi + f\phi' - (\tfrac{1}{2}P_r^{-1}\theta_2 + f'^2)\phi/a = 0 \tag{9a}$$

$$\phi'(0) = 0, \quad \phi(\infty) = \eta - \delta \tag{9b,c}$$

The farfield condition (9c) is imposed in its differentiated form $\phi'(\infty) = 1$.

To fix the problem for the numerical treatment, we choose $UL = \nu$ which renders $R_e = 1$ and $R_m = \sigma\mu\nu \equiv P_m$, the magnetic Prandtl number. The further choice $a = 1$ corresponds to $U/L$ measuring the strength of the SPF. Then $b$ is the ratio of the stretching rate to the SPF strength.

## 5. Results and discussion

The problem involves as parameters, the magnetic Prandtl number $P_m$, the magnetic interaction number $\beta$, the stretching ratio $b$, and the Prandtl number $P_r$.

In tables 1-3, we present results for the surface shear $f''(0)$, the surface components of the induced magnetic field $g'(0)$ and $g(0)$, and the constituents of the surface temperature $\theta_0(0)$ and $\theta_2(0)$, corresponding to $P_r = 0.72$ and wide ranges of $P_m$, $\beta$ and $b$, respectively.

As $P_m \sim 0$, there is tendency toward a limit in which $g \sim \eta - \delta$, in the leading order. The induced magnetic field gets negligibly smaller. The imposed magnetic field, which is parallel to the farfield velocity, progressively extends everywhere. This is in line with the case of the flow being exclusively due to a linearly stretching sheet [11].

As $\beta \sim 0$, the velocity and temperature fields are not affected by the Lorentz force and Joule heating, in the leading order. The magnetic field, on the other hand, develops an induced part.

In the critical case when the stretching rate and the SPF strength are equivalent ($b=1$), the problem has the exact solution $f = a\eta$, $g = \eta$ and $\theta_2 = 0$, while $\theta_0$ is zero only when viscous dissipation and Joule heating are neglected. As $b$ deviates from unity, the surface values of $f''$, $g'$, $g$ and $\theta_2$ progressively deviate from their corresponding critical values. The rise in surface shear $|f''(0)|$ indicates higher dissipation which is manifested in the rise of $\theta_2(0)$. The initial drop in $\theta_0(0)$, as $b$ falls below unity, is due to the drop in the dissipation term $f'^2(0) = b^2$ which is not compensated for by the small rise in $\theta_2(0)$.

Profiles of the velocity and magnetic field components are depicted in Figs. 1-3, for the case of a pure SPF ($b=0$) and cases of SPF toward a stretching sheet ($b=0.5$, 1 and 2.0). Stretching the sheet results in increase in the streamwise speed $f'(\eta)$, which is compensated for by increase in the speed toward the sheet $f(\eta)$. These increases are responsible for the increase of the magnetic field components $g(\eta)$ and $g'(\eta)$.

Profiles of the temperature constituents $\theta_0(\eta)$ and $\theta_2(\eta)$ are shown in Figs. 4 and 5, at two values of the Prandtl number- $P_r = 0.72$ for air and $P_r = 7.0$ for water, around the room conditions. The higher the Prandtl number the higher the surface temperature. There is more accumulation of heat close to the insulated surface due to the lower thermal diffusivity associated with higher $P_r$. The faster tendency toward the farfield limit, at higher $P_r$, indicates the formation of a thermal boundary layer.

Table 1: Effect of varying $P_m$; $\beta = 0.1$, $b = 0$, $P_r = 0.72$

| $P_m$ | $f''(0)$ | $g'(0)$ | $g(0)$ | $\theta_0(0)$ | $\theta_2(0)$ |
|---|---|---|---|---|---|
| $10^{-1}$ | 1.26589 | 1.02987 | -0.658859 | 2.28996 | 0.437047 |
| $10^{-2}$ | 1.26399 | 1.00284 | -0.633616 | 2.28737 | 0.436243 |
| $10^{-3}$ | 1.26381 | 1.00028 | -0.631233 | 2.28713 | 0.436169 |
| $10^{-4}$ | 1.26379 | 1.00003 | -0.630996 | 2.28711 | 0.436161 |
| $10^{-5}$ | 1.26379 | 1.00000 | -0.630972 | 2.28710 | 0.436160 |

Table 2: Effect of varying $\beta$; $P_m = 0.1$, $b = 0$, $P_r = 0.72$

| $\beta$ | $f''(0)$ | $g'(0)$ | $g(0)$ | $\theta_0(0)$ | $\theta_2(0)$ |
|---|---|---|---|---|---|
| $10^{-2}$ | 1.23618 | 1.03168 | -0.677558 | 2.2523 | 0.425252 |
| $10^{-1}$ | 1.26589 | 1.02987 | -0.658859 | 2.2900 | 0.437047 |
| 1 | 1.45429 | 1.02162 | -0.563371 | 2.5046 | 0.510809 |
| 10 | 2.06882 | 1.01015 | -0.387526 | 3.0482 | 0.743698 |
| $10^2$ | 3.38990 | 1.00369 | -0.233921 | 3.9106 | 1.224236 |

Table 3: Effect of varying $b$; $P_m=0.1$, $\beta=0.1$, $P_r=0.72$

| $b$ | $f''(0)$ | $g'(0)$ | $g(0)$ | $\theta_0(0)$ | $\theta_2(0)$ |
|---|---|---|---|---|---|
| 0 | 1.26589 | 1.02987 | -0.658859 | 2.2900 | 0.437047 |
| 0.2 | 1.07907 | 1.02824 | -0.504150 | 1.7492 | 0.260974 |
| 0.4 | 0.85624 | 1.02391 | -0.363205 | 1.5272 | 0.139408 |
| 0.6 | 0.60053 | 1.01749 | -0.233401 | 1.5136 | 0.059535 |
| 0.8 | 0.31442 | 1.00941 | -0.112816 | 1.6454 | 0.014419 |
| 1.0 | 0.000000 | 1.000000 | 0.000000 | 1.883817 | 0.000000 |
| 1.2 | -0.341013 | 0.989482 | 0.106171 | 2.203459 | 0.013756 |
| 1.4 | -0.707141 | 0.978040 | 0.206585 | 2.586852 | 0.054044 |
| 1.6 | -1.097109 | 0.965817 | 0.301960 | 3.021518 | 0.119758 |
| 1.8 | -1.509805 | 0.952923 | 0.392882 | 3.498253 | 0.210124 |
| 2.0 | -1.944247 | 0.939450 | 0.479836 | 4.010094 | 0.324591 |

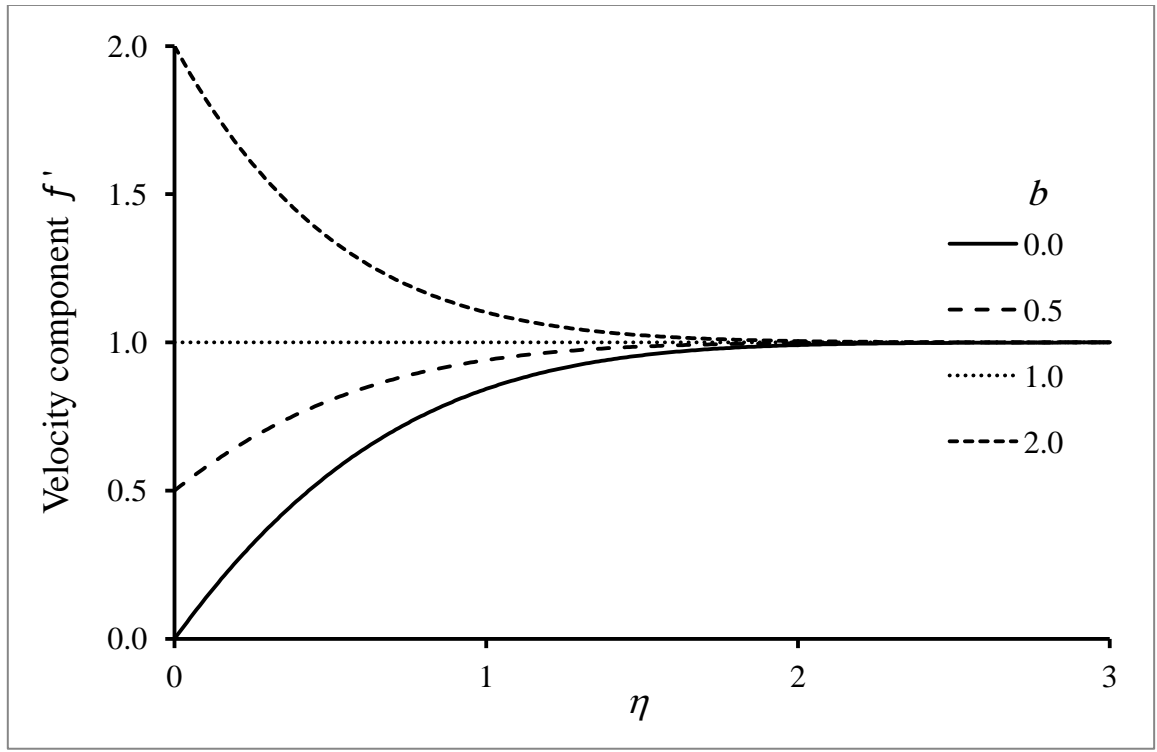

Fig. 1: Profiles of the streamwise velocity component; $P_m=0.1$, $\beta=1.0$.

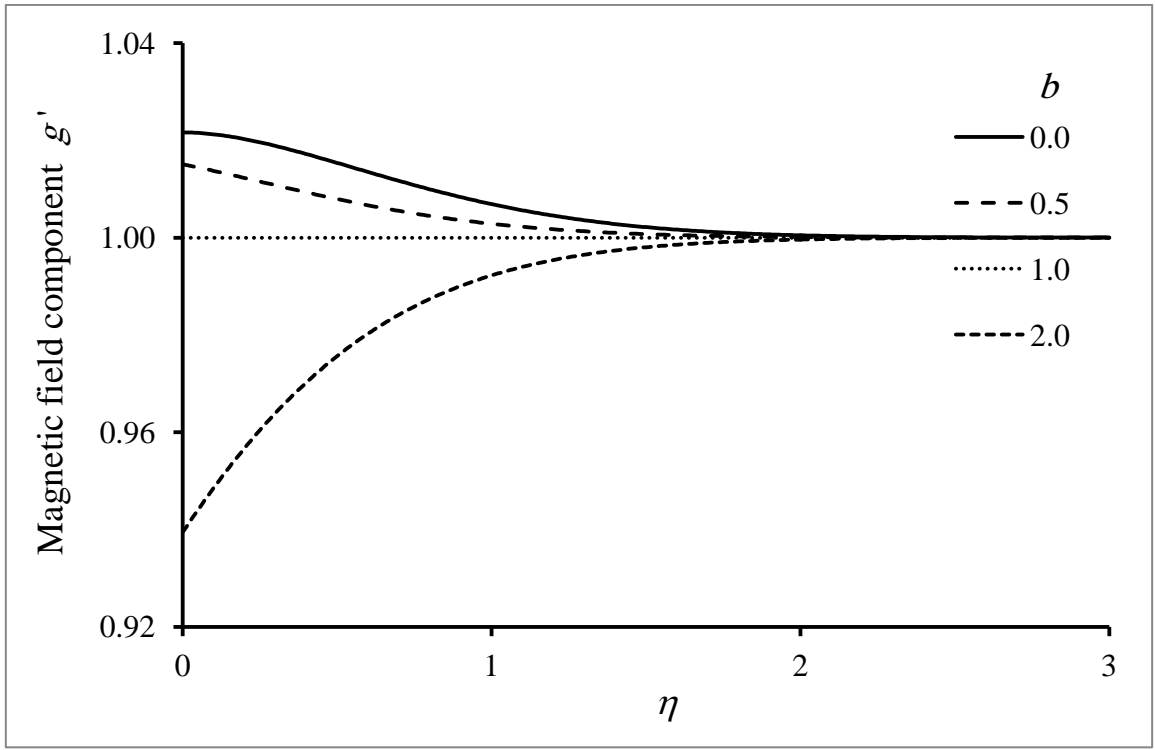

Fig. 2: Profiles of the streamwise magnetic field component; $P_m$ =0.1, $\beta$ =1.0.

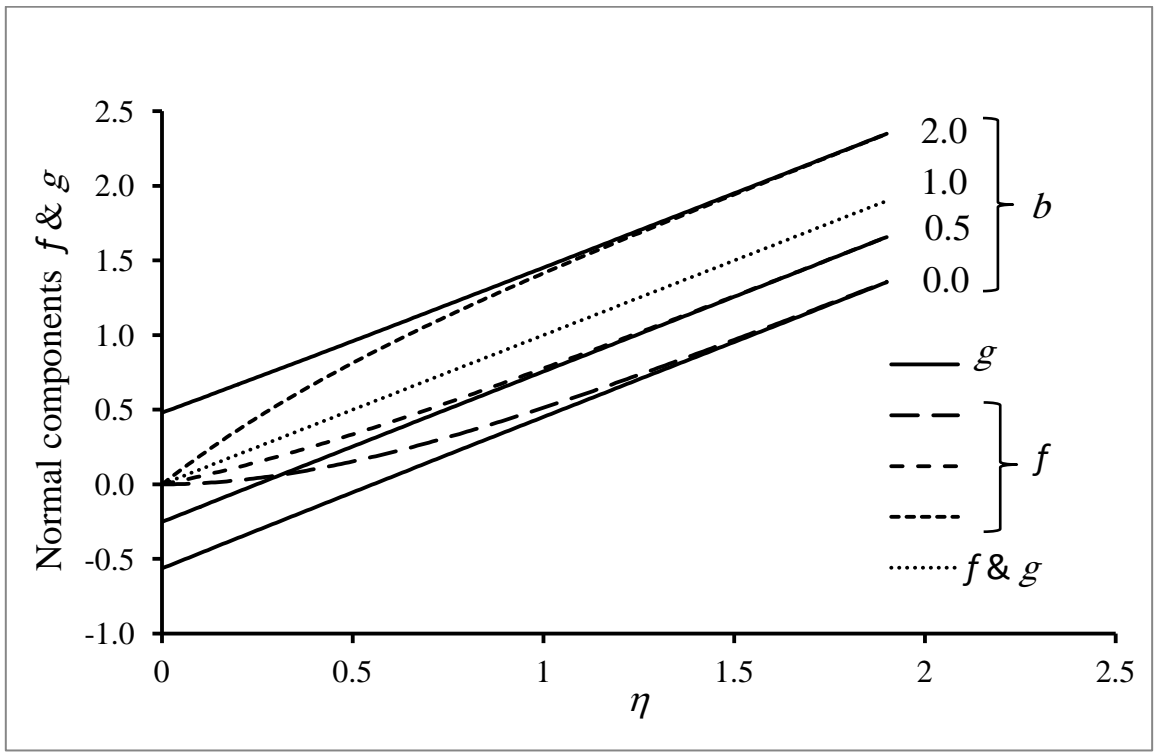

Fig. 3: Profiles of the normal velocity and magnetic field components; $P_m$ =0.1, $\beta$ =1.0.

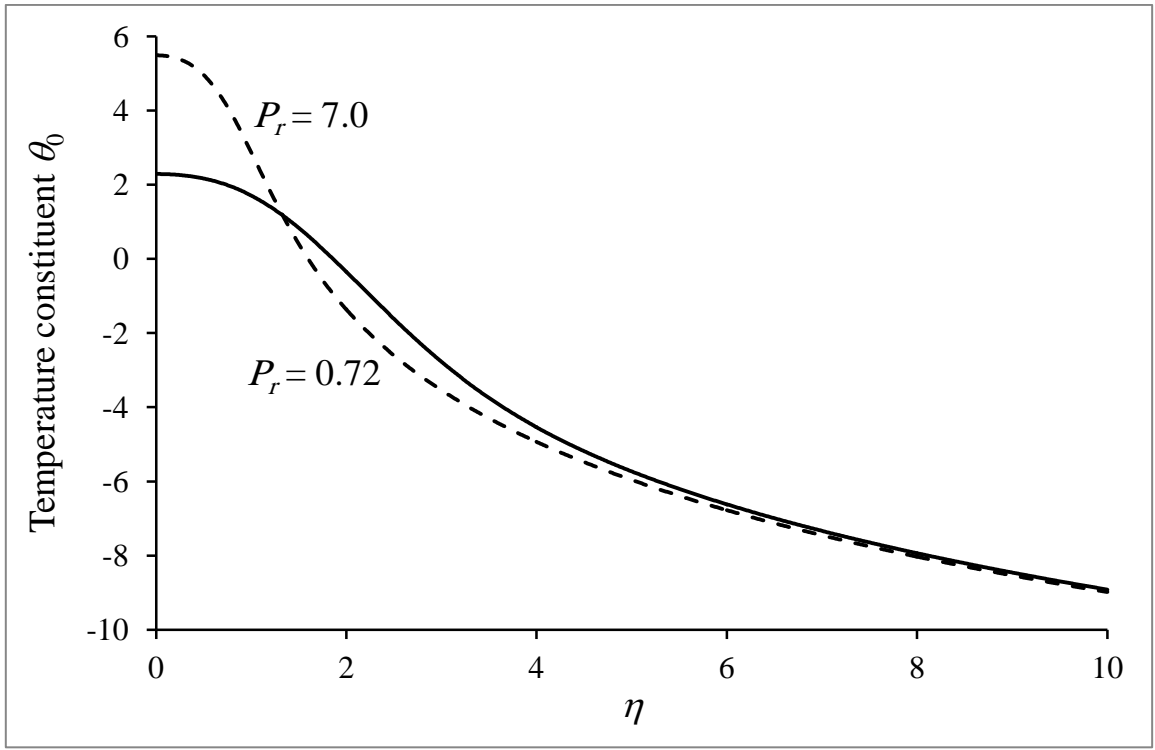

Fig. 4: Profiles of the temperature constituent $\theta_0$; $P_m = 0.1$, $\beta = 1.0$, $b = 0.0$.

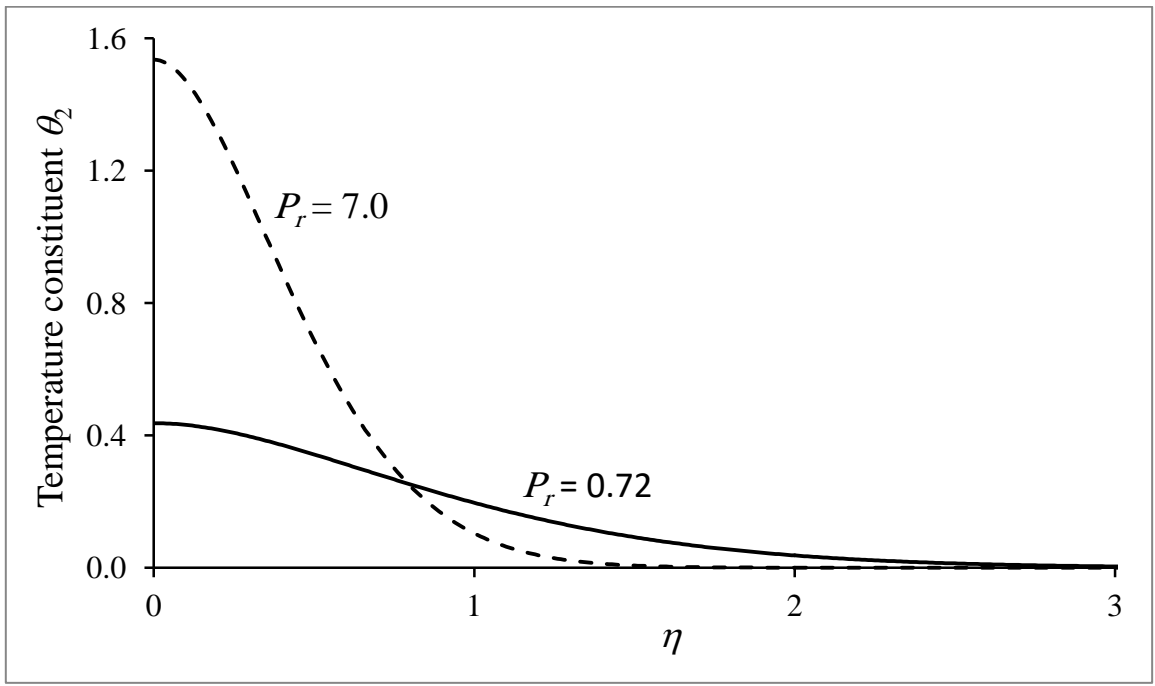

Fig. 5: Profiles of the temperature constituent $\theta_2$; $P_m = 0.1$, $\beta = 1.0$, $b = 0.0$.

## 6. Conclusion

The MHD problem of the SPF flow toward a linearly stretching sheet is shown to admit self similarity of the full governing fluid flow and electromagnetic equations, provided that the farfield velocity and magnetic fields are parallel.

Viscous dissipation and Joule heating are included in the energy equation, and the case of a thermally insulated surface is studied. The viscous dissipation introduces a correction to the farfield temperature, an expression for which is obtained. The temperature decomposes into a streamwise non-evolving part, and a quadratically evolving part .

Four parameters are involved, the magnetic Prandtl number $P_m$, the magnetic interaction number $\beta$, the stretching/SPF ratio $b$ and the Prandtl number $P_r$. Numerical solutions are obtained for wide ranges of these parameters and exhibited through tables and figures.

**Funding**: This research did not receive any specific grant from funding agencies in the public, commercial, or not-for-profit sectors.